\documentclass[aps,superscriptaddress,showpacs,dvips]{revtex4}
\usepackage{feynmp}
\usepackage{amssymb}
\usepackage{epsfig}
\usepackage{graphicx}
\usepackage{subfigure}
\usepackage{hyperref}

\begin{document}

\title{Eliashberg theory of excitonic insulating transition in graphene}

\author{Jing-Rong Wang}
\affiliation{Department of Modern Physics, University of Science and
Technology of China, Hefei, Anhui 230026, P. R. China}
\author{Guo-Zhu Liu}
\affiliation{Department of Modern Physics, University of Science and
Technology of China, Hefei, Anhui 230026, P. R. China}
\affiliation{Institut f$\ddot{u}$r Theoretische Physik, Freie
Universit$\ddot{a}$t Berlin, Arnimallee 14, D-14195 Berlin, Germany}

\begin{abstract}
A sufficiently strong Coulomb interaction may open an excitonic
fermion gap and thus drive a semimetal-insulator transition in
graphene. In this paper, we study the Eliashberg theory of excitonic
transition by coupling the fermion gap equation self-consistently to
the equation of vacuum polarization function. Including the fermion
gap into polarization function increases the effective strength of
Coulomb interaction because it reduces the screening effects due to
the collective particle-hole excitations. Although this procedure
does not change the critical point, it leads to a significant
enhancement of the dynamical fermion gap in the excitonic insulating
phase. The validity of the Eliashberg theory is justified by showing
that the vertex corrections are suppressed at large $N$ limit.
\end{abstract}


\maketitle


The low-energy elementary excitations of graphene are known to be
massless Dirac fermions with linear dispersion \cite{CastroNeto,
DasSarma}. When Dirac fermions are strictly massless, the system has
a semimetal ground state and its Hamiltonian possesses a continuous
chiral symmetry. However, when the Coulomb interaction is
sufficiently strong, the massless fermion (particle) may combine
with an anti-fermion (hole) to form a stable excitonic pair
\cite{Khveshchenko, Gusynin}. As a consequence, the Dirac fermion
acquires a finite mass gap, which dynamically breaks the chiral
symmetry, and the system undergoes a semimetal-insulator transition.
In recent years, this excitonic transition has been investigated by
several tools, including Dyson-Schwinger gap equation
\cite{Khveshchenko, Gusynin, Liu, LiuWang, Gamayun}, Monte Carlo
simulation \cite{Hands08, Hands10, Drut09_1, Drut09_2}, and
renormalization group \cite{Herbut}. In most of these research
works, it was found that such transition can take place when Coulomb
interaction strength parameter $\lambda$ is larger than certain
critical value $\lambda_c$. The opening of fermion gap can result in
important changes in the low-energy properties \cite{Gusynin,
Gusynin06, Kotov, Liupla, CastroNetoPhys}.

Currently, there is still no compelling experimental evidence for
the existence of excitonic insulating transition in graphene. In
general, there are three possible reasons why the excitonic
transition has not yet been unambiguously observed in experiments.
First, the Coulomb interaction may be too weak to induce excitonic
pair formation. Second, the Coulomb interaction is sufficiently
strong in certain graphene materials, but the excitonic transition
is suppressed by various fluctuations and perturbations. Actually,
we have recently examined this possibility and showed \cite{Liu,
LiuWang} that even for sufficiently strong Coulomb interaction the
excitonic transition can be destroyed by thermal fluctuations,
finite doping, disorder scattering, and finite lattice effect.
Third, the excitonic transition does happen in some graphene, but
the fermion gap is too small in magnitude to be clearly resolved by
any experimental instruments. At present, it is not possible to
judge what exactly the reason is. In order to examine the third
possibility, it is necessary to determine the magnitude of fermion
gap precisely.

Most previous research efforts focused on an accurate determination
of the critical point $\lambda_c$ \cite{Khveshchenko, Gusynin, Liu,
LiuWang, Gamayun, Hands08, Hands10, Drut09_1, Drut09_2,
CastroNetoPhys}. However, an accurate determination of dynamical
fermion gap should be equally important. As discussed above, the
predicted excitonic insulator is experimentally detectable only when
the fermion gap is sufficiently large. On the other hand, the
graphene with a large gap will have many technological advantages
\cite{CastroNetoPhys}. As emphasized by Castro Neto
\cite{CastroNetoPhys}, if the fermion gap is too small, the
interests of excitonic insulating transition would be purely
academic. Therefore, an exact determination of the dynamical fermion
gap is very important both experimentally and technologically.

In most previous gap equation analysis of the excitonic transition,
the vacuum polarization function appearing in the Coulomb
interaction function was calculated using the free propagator of
massless Dirac fermion, which amounts to assuming the random phase
approximation (RPA). Thus the feedback effect of the dynamical
fermion gap on the Coulomb interaction was simply ignored. The
importance of such feedback effect can be seen by making a simple
qualitative analysis. As shown in previous works \cite{Liu,
LiuWang}, the screening of Coulomb interaction due to the collective
particle-hole excitations can have very important influence on the
excitonic transition because it reduces the effective interaction
strength. Technically, the screening of Coulomb interaction is
described by the polarization function $\Pi(\mathbf{q})$. At the
level of RPA, $\Pi(\mathbf{q}) \propto |\mathbf{q}|$, so the Coulomb
interaction is weakened due to dynamical screening. Once a finite
fermion gap $m$ is included in the polarization function, we have
$\Pi(\mathbf{q}) \propto \mathbf{q}^2/m$. This term is much smaller
than the term $\propto |\mathbf{q}|$ in the low momentum regime, so
the effective Coulomb interaction becomes much stronger. From this
qualitative analysis, we know that the feedback effect of the
dynamical fermion gap may play an important role in the gap equation
analysis of the excitonic transition.

In order to compute the dynamical fermion gap more accurately, it is
crucial to study the equations of fermion gap and polarization
function in a self-consistent manner. This formalism corresponds to
the Eliashberg theory \cite{Carbotte}, which was originally
developed to describe the unusual properties of conventional
superconductors with strong electron-phonon coupling
\cite{Carbotte}. Apart from its remarkable success in studying
electron-phonon interaction induced superconductors \cite{Carbotte},
the Eliashberg theory is also useful in many other condensed matter
problems. In particular, it is widely adopted when studying the
non-Fermi liquid behaviors in some correlated electron systems with
singular fermion-boson interactions \cite{Polchinski, WangLiu,
Vojta, Paaske, Chubukov04, Chubukov06, Chubukov10}.

In this paper, we study the Eliashberg theory of the excitonic gap
generation in graphene. To make a general analysis, we include both
finite temperature and finite chemical potential. After solving the
self-consistent equations of fermion gap function and polarization
function, we found that the critical point does not change, which is
reasonable because bifurcation theory ensures that the fermion gap
can be safely taken to be zero near the critical point. However, the
size of dynamical fermion gap is significantly enhanced compared to
that obtained using the RPA polarization function. This implies that
the self-consistent Eliashberg theory plays a crucial role in an
accurate computation of dynamical fermion gap. The Eliashberg theory
is justified only when the vertex corrections are unimportant.
Within the $1/N$ expansion, we show that the vertex corrections are
suppressed in the large $N$ limit.

The Hamiltonian for interacting Dirac fermions is
\begin{eqnarray}
H &=& v_{F}\sum_{i=1}^{N}\int_{\mathbf{r}}
\bar{\psi}_{i}(\mathbf{r})i\mathbf{\gamma}\cdot\mathbf{\nabla}
\psi_{i}(\mathbf{r}) \nonumber \\
&& + \frac{1}{4\pi}\sum_{i,j}^{N}
\int_{\mathbf{r},\mathbf{r}^{\prime}}
\bar{\psi}_{i}(\mathbf{r})\gamma_{0}\psi_{i}(\mathbf{r})
\frac{e^{2}}{|\mathbf{r}-\mathbf{r}^{\prime}|}
\bar{\psi}_{j}(\mathbf{r}^{\prime})\gamma_{0}
\psi_{j}(\mathbf{r}^{\prime}).
\end{eqnarray}
As usual, we adopt the four-component spinor field $\psi$ to
describe Dirac fermion and define the conjugate spinor field as
$\bar{\psi} = \psi^{\dagger} \gamma_{0}$. The $4\times 4$
$\gamma$-matrices satisfy the Clifford algebra \cite{Khveshchenko,
Gusynin}. The physical flavor of Dirac fermion is $N=2$. The total
Hamiltonian possesses a continuous chiral symmetry $\psi \rightarrow
e^{i\theta\gamma_5}\psi$, where $\gamma_5$ anticommutes with
$\gamma_{\mu}$, which will be dynamically broken once a nonzero
fermion gap is generated.

The free propagator for massless Dirac fermion is
$G_{0}^{-1}(p_{0},\mathbf{p}) = \gamma_{0} p_{0} -
v_{F}\mathbf{\gamma}\cdot \mathbf{p}$. After including the
interaction effect, it is modified to
\begin{eqnarray}
G^{-1}(p_{0},\mathbf{p}) = A_{0}p_{0}\gamma_{0} - v_{F}A_1
\mathbf{\gamma}\cdot\mathbf{p} - m(p_{0},\mathbf{p}),
\end{eqnarray}
where $A_{0,1}(p_{0},\mathbf{p})$ is the wave function
renormalization and $m(p_{0},\mathbf{p})$ is the fermion gap
function. These quantities can in principle be obtained by solving
the following complete Dyson-Schwinger equation
\begin{eqnarray}
G^{-1}(p) &=& G_{0}^{-1}(p) + \int \frac{d^{3}k}{(2\pi)^{3}}
\gamma_{0}G(k)\Gamma_{0}V(p-k),
\end{eqnarray}
where $\Gamma_{0}$ is the vertex function and $V(q)$ is the Coulomb
interaction function. The bare Coulomb interaction is $V_{0}(q) =
\frac{e^2}{2\epsilon_{0}v_{F}|\mathbf{q}|}$, which is clearly
long-ranged. After taking into account dynamical screening effect
from collective particle-hole excitations, the effective Coulomb
interaction function is modified to
\begin{equation}
V(q) = \frac{1}{V_{0}^{-1}(q)+\Pi(q)},
\end{equation}
with $\Pi(q)$ being the vacuum polarization function. The above gap
equation has been studied extensively in recent years and dynamical
gap generation was found when Coulomb interaction is sufficiently
strong \cite{Khveshchenko, Gusynin, Liu, LiuWang, Gamayun}. In most
of these treatments, only the leading order of $1/N$ expansion was
kept. In particular, both wave function renormalzation and vertex
corrections were neglected \cite{Khveshchenko, Gusynin, Liu,
LiuWang, Gamayun}, so that $A_{0,1} = 1$ and $\Gamma_{0} =
\gamma_{0}$. Moreover, the massless fermion propagator is widely
used when calculating the polarization function shown in Fig.1 and
the feedback effect of fermion mass is simply ignored. Within this
approximation, the polarization function has the form
\begin{equation}
\Pi(q) = \frac{N}{8}\frac{\mathbf{q}^{2}}{\sqrt{q_{0}^2 +
v_{F}^{2}|\mathbf{q}|^2}}.
\end{equation}

\begin{figure}
\center
\includegraphics[width=1.98in]{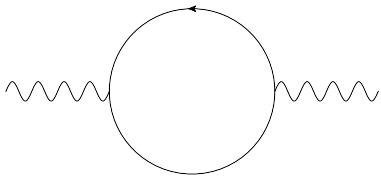}
\caption{The bubble Feynman diagram for the polarization function.
The solid line is the free fermion propagator and the wavy line is
the Coulomb interaction function.}\label{polarization}
\end{figure}

The aim of this work is to go beyond the popular RPA calculation and
include the dynamical fermion gap back into the polarization
function shown in Fig.1. The importance of this treatment can be
readily seen by making a simple qualitative analysis. Generally, the
effective strength of Coulomb interaction is characterized by two
ingredients: interaction parameter $\lambda =
e^2N/16v_{F}\epsilon_0$ with $\epsilon_0$ being the dielectric
constant and static/dynamical screening due to collective
particle-hole excitations. While the former is determined by the
substrate of graphene sample, the latter is reflected in the
polarization function. Within RPA, the polarization function (5)
behaves as $\propto \mathbf{q}$ in the static limit $q_{0} = 0$. It
vanishes linearly as $\mathbf{q}\rightarrow 0$, so the Coulomb
interaction remains long-ranged. However, the effective Coulomb
interaction is weakened by the dynamical screening term $\propto
\mathbf{q}$, which arises from particle-hole excitations. In the
chiral symmetry broken phase, the Dirac fermion has a finite mass.
Assuming a constant mass $m$, the polarization function becomes
\begin{eqnarray}
\Pi(q) = \frac{N}{\pi}\mathbf{q}^{2}\left(\frac{m}{2q^{2}} +
\frac{q^{2} - 4m^{2}}{4q^3} \arcsin\frac{q}{\sqrt{q^2+4m^2}}\right).
\end{eqnarray}
In the limit $q \ll m$, it is simplified to
\begin{eqnarray}
\Pi(\mathbf{q}) \propto \frac{\mathbf{q}^{2}}{m}.
\end{eqnarray}
At the low energy regime, this contribution is much less than the
term $\propto \mathbf{q}$ of the polarization function obtained
using the massless fermion propagator (RPA). Physically, this
reflects the fact that a finite fermion gap reduces the dynamical
screening effect. Because of this reduction, the effective Coulomb
interaction becomes much stronger.

In general, the fermion mass gap is not a constant, but depends
explicitly on momentum. To examine the feedback effect of fermion
gap on excitonic transition, we will utilize the Eliashberg
formalism and couple the fermion gap equation to the equation of
polarization function self-consistently.

\begin{figure}[h]
\vspace{-3.5cm} \center
   \includegraphics[width=2.5in]{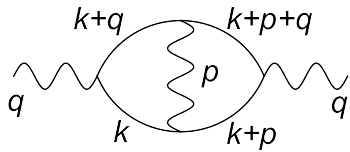}
\caption{The polarization operator with the vertex correction, but
without self-energy correction to intermediate fermions.}
     \label{fig:SP}
\end{figure}

In the Eliashberg formalism, the vertex corrections are usually
ignored. This approximation is well-justified in the electron-phonon
interacting systems as the Migdal theorem ensures that the vertex
corrections are suppressed by a small factor $m/M$, where $m$ is the
electron mass and $M$ is the nuclei mass \cite{Carbotte}. In the
present problem, we still have a small suppressing factor in the
vortex corrections based on the $1/N$ expansion. This can be
explained by considering the vertex correction diagram for
polarization shown in Fig.2. In the large $N$ limit, it is found
(details are presented in the Appendix) to have the form
\begin{eqnarray}
\Pi_{\mathrm{v}}(q) = -\frac{8}{\pi^2}
\frac{\ln\left(N\frac{e^2}{8\epsilon_0 v_{F}}\right)
\ln\left(\frac{\Lambda}{q}\right)}{N}\Pi (q).
\end{eqnarray}
Apparently, we know that
\begin{eqnarray}
\frac{\Pi_{\mathrm{v}}(q)}{\Pi (q)} &\propto&
\frac{\ln\left(N\frac{e^2}{8\epsilon_{0}v_{F}}\right)}{N},
\end{eqnarray}
so the polarization function with vertex correction
$\Pi_{\mathrm{v}}(q)$ is suppressed by a small factor $\ln(N)/N$
comparing with the leading polarization function $\Pi(q)$, and
therefore can be neglected in the large $N$ limit. One can verify in
a similar way that the same is also true with the vertex correction
diagram for the fermion self-energy. Similar arguments for the
suppression of vertex corrections are extensively used in the
Eliashberg theories of a plenty of physical problems, including
fermion-gauge systems \cite{Polchinski, WangLiu}, quantum critical
points in itinerant electron systems \cite{Vojta, Paaske,
Chubukov04, Chubukov06}, and electron-doped cuprate superconductors
\cite{Chubukov10}. Moreover, the wave function renormalization
$A_{0,1}(p_{0},\mathbf{p})$ also contain certain powers of $1/N$.
Therefore, in this paper we will ignore both vertex corrections and
wave function renormalizations.

In order to make a more general analysis, we consider the problem at
finite temperature $T$ and finite chemical potential $\mu$ and study
how the static screening of Coulomb interaction due to finite $T$
and $\mu$ is affected by fermion mass. We will work in the Matsubara
formalism and replace the fermion energy $p_{0}$ by an imaginary
frequency, $p_{0} \rightarrow i\omega_{n}=i(2n+1)\pi T$. At finite
$T$, it is convenient to adopt an instantaneous approximation
\cite{Khveshchenko, Gusynin} and ignore the energy-dependence of the
interaction function. At finite $\mu$, the frequency $i\omega_n$
appearing in fermion propagator should be replaced by $i\omega_n -
\mu$. The instantaneous approximation allows us to perform the
frequency summation over $k_{0}$ analytically, leading to the gap
equation
\begin{eqnarray}
m(\mathbf{p}) &=& \frac{1}{4N}\sum_{\alpha=\pm1}\int
\frac{d^{2}\mathbf{k}}{(2\pi)^{2}}
\frac{m(\mathbf{k})}{\sqrt{\mathbf{k}^2 + m^2(\mathbf{k})}}
\nonumber \\
&& \times \tanh\Big(\frac{\sqrt{\mathbf{k}^2 +
m^2(\mathbf{k})}+\alpha \mu}{2T}\Big)
\frac{1}{\frac{\mathbf{|q|}}{8\lambda} +
\frac{1}{N}\Pi(\mathbf{q},T)},
\end{eqnarray}
where $\mathbf{q} = \mathbf{p-k}$. We need to compute the
polarization function $\Pi(\mathbf{q},T)$ that contains the
dynamical fermion mass function $m(\mathbf{q})$. Within the
instantaneous approximation, the polarization function is defined as
\begin{eqnarray}
\Pi(\mathbf{q}) = -\frac{N}{\beta}
\sum_{n}\int\frac{d^{2}\mathbf{k}}{(2\pi)^{2}}\mathrm{Tr}
\left[G(\omega_{n},\mathbf{k}) \gamma_{0}
G(\omega_{n},\mathbf{k}+\mathbf{q})\gamma_{0}\right], \nonumber
\end{eqnarray}
where the Dirac fermion propagator $G(\omega_{n},\mathbf{k})$
contains dynamical fermion mass $m(\mathbf{p})$. Following the
procedure presented in a previous work \cite{Liu}, we can sum over
imaginary frequency $\omega_{n}$ and then obtain
\begin{eqnarray}
\Pi(\mathbf{q},T) &=& \frac{N}{2}\sum_{\alpha=\pm1}\int_{0}^{1}dx
\int\frac{d^{2}\mathbf{k}}{(2\pi)^{2}}\frac{1}{t^2} \nonumber \\
&& \times \left[\frac{t'}{t} \tanh(\frac{t+\alpha\mu}{2T}) +
\frac{t''}{2T}\frac{1}{\cosh^2(\frac{t+\alpha\mu}{2 T})}\right],
\end{eqnarray}
where we defined three parameters
\begin{eqnarray}
t &=& [\mathbf{k}^2+x^2\mathbf{q}^2 + 2x\mathbf{k\cdot
q}+(1-x)m^{2}(\mathbf{k}) \nonumber \\
&& +xm^2(\mathbf{k+q}) + x(1-x)\mathbf{q}^{2}]^{1/2}, \nonumber \\
t' &=& x \mathbf{q}^2 + (2x-1)\mathbf{k\cdot q} +
(1-x)m^{2}(\mathbf{k}) \nonumber \\
&& + x m^2(\mathbf{k+q}) - m(\mathbf{k})m(\mathbf{k+q}), \nonumber
\\
t'' &=& 2 \mathbf{k}^2 + x \mathbf{q}^2 + (2x+1)\mathbf{k\cdot q}
+ (1-x)m^{2}(\mathbf{k}) \nonumber \\
&& + xm^2(\mathbf{k+q}) + m(\mathbf{k})m(\mathbf{k+q}). \nonumber
\end{eqnarray}
Now we obtain the self-consistent equations (10) and (11) for
dynamical fermion mass function and polarization function. The
dynamical fermion mass can be obtained by solving them numerically.

Here, we would like to make some remarks on the instantaneous
approximation. This approximation was originally proposed in the
study of dynamical fermion gap generation at finite temperature in
QED$_3$ \cite{Doreypl, Doreynp}. Technically, when calculating the
fermion gap equation and polarization function, it is possible to
sum over the imaginary frequencies in the Matsubara formalism only
when the energy (frequency) dependence of polarization function is
neglected. Otherwise, there will be infinitely many coupled gap
equations \cite{Doreypl, Doreynp}. Moreover, in order to make a
qualitative analysis of the screening effects, we need to obtain an
(semi)analytical expression for the polarization function, as
Eq.(11), which can be derived only within the instantaneous
approximation. At zero temperature and zero chemical potential, it
is formally viable to include the energy dependence of the
polarization function. Indeed, we have solved the fermion gap
equation without assuming the instantaneous approximation
\cite{Liu}, but using the RPA polarization function (5). The effects
of the dynamical part of the RPA polarization function was also
investigated in a recent paper of Gamayun \emph{et} \emph{al.}
\cite{Gamayun}. Due to the non-relativistic nature of the present
Coulomb-interacting system, the integrations over energy and momenta
have to be performed separately, which substantially increases the
time needed to perform numerical computations. In the Eliashberg
formalism, we need to go beyond the RPA level and solve the
self-consistently coupled equations for the fermion gap and the
polarization function. This requires much more computer time and
would significantly reduce the precision of numerical output. In
order to retain the necessary numerical precision, we use the
instantaneous approximation even at zero temperature and zero
chemical potential.

\begin{figure}[h]
\center
   \includegraphics[width=2.94in]{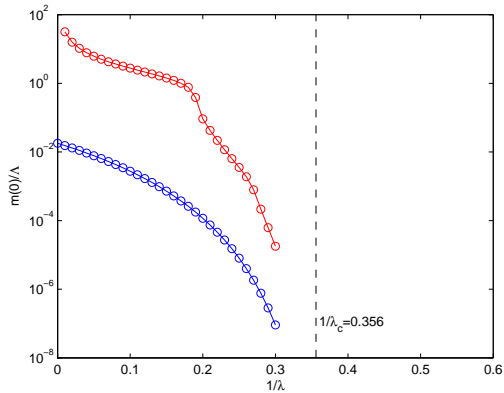}
\caption{The $\lambda$ dependence of dynamical fermion gap at zero
momentum and zero temperature. At physical flavor $N=2$, the
critical strength is given by $1/\lambda_c = 0.356$. When $\lambda <
\lambda_c$, the excitonic insulating transition can not happen.}
\label{fig:GapPSF}
\end{figure}

\begin{figure}[h]
\center
\includegraphics[width=2.89in]{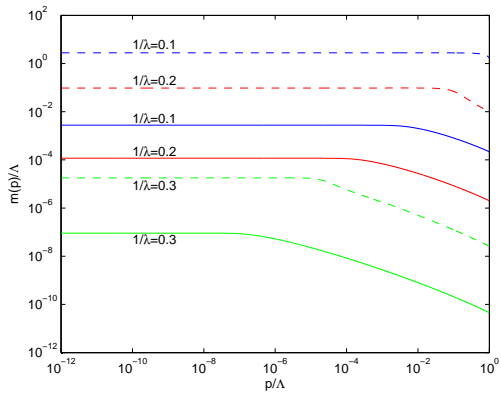}
\caption{Dynamical fermion gap at zero temperature.}
\label{fig:GapZeroT}
\end{figure}

Before performing numerical computation, it is helpful to first
qualitatively analyze the effect of finite fermion gap on the static
screening of Coulomb interaction. As an example, we consider the
case of finite chemical potential at zero temperature. At $T=0$, the
fermion gap equation is simplified to
\begin{eqnarray}
m(\mathbf{p}) = \frac{1}{N}\int\frac{d^{2}\mathbf{k}}{8\pi^{2}}
\frac{m(\mathbf{k})}{\sqrt{\mathbf{k}^2 + m^2(\mathbf{k})}}
\frac{\theta(\sqrt{\mathbf{k}^2 +
m^2(\mathbf{k})}-\mu)}{\frac{\mathbf{|p-k|}}{8\lambda} +
\frac{1}{N}\Pi(\mathbf{p-k})}, \nonumber
\end{eqnarray}
which couples to the polarization function
\begin{eqnarray}
\Pi(\mathbf{q}) = N\int_{0}^{1}dx\int
\frac{d^{2}\mathbf{k}}{(2\pi)^{2}}\left[\frac{t'}{t^3}\theta(t-\mu)
+\frac{t''}{t^2}\delta(t-\mu)\right]. \nonumber
\end{eqnarray}
Now we assume a constant fermion mass gap $m$, then the integration
over momentum $\mathbf{k}$ in these equations can be carried out
exactly. When $\mu > \sqrt{m^{2} + \mathbf{q}^2/4}$, the
polarization function behaves as $\Pi(\mathbf{q}) = N\mu/\pi$. This
expression implies that the Coulomb interaction is now statically
screened by finite chemical potential and thus becomes short-ranged.
Such static screening effect will rapidly destroy the excitonic
pairing instability \cite{Liu}. However, when the fermion gap is
relatively large, $m > \mu$, we have $\Pi(\mathbf{q})\propto
\mathbf{q}^2/m$ in the low momentum regime, so the Coulomb
interaction remains long-ranged and is only poorly screened. From
this qualitative analysis, we know that once the feedback effect of
fermion gap is included, the static screening may be suppressed and
the effective Coulomb interaction may still be strong even at finite
chemical potential. Besides chemical potential, thermal fluctuation
can also induce static screening \cite{Liu}. The effect of fermion
gap on static screening at finite temperature can be qualitatively
analyzed similarly. To gain more quantitative understanding on
feedback effect of fermion gap, we have to perform numerical
computation.

\begin{figure}[h]
\center
\includegraphics[width=2.89in]{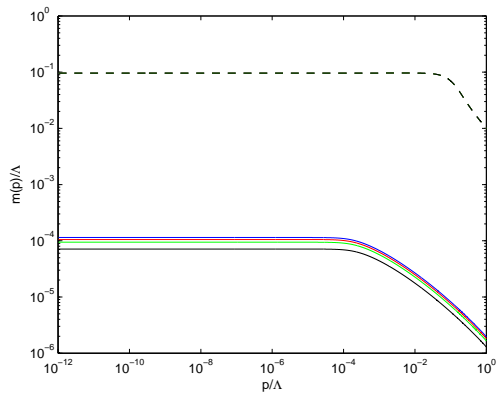}
\caption{Dynamical fermion gap at finite temperature and zero
chemical potential, with $1/\lambda = 0.2$.} \label{fig:GapFinitT}
\end{figure}

\begin{figure}[h]
\center
\includegraphics[width=2.89in]{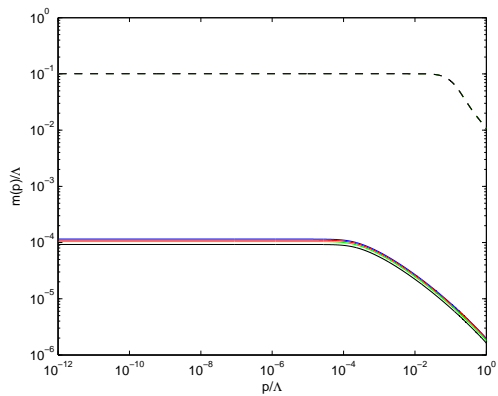}
\caption{Dynamical fermion gap at finite chemical potential with
$T/\Lambda = 10^{-7}$ and $1/\lambda = 0.2$.} \label{fig:GapFinitMu}
\end{figure}

We have numerically solved the coupled equations of dynamical
fermion gap and polarization function by means of straightforward
iterative method. When $T=0$, the fermion gap of zero momentum is
shown in Fig.\ref{fig:GapPSF}, where the blue (red) curve represents
the results obtained using the perturbative (self-consistent)
polarization function. From the numerical results, we know that the
critical Coulomb interaction parameter $\lambda_c$ takes the same
value in these two cases ($1/\lambda_c = 0.356$ for physical flavor
$N=2$). This fact is easy to understand since the bifurcation theory
requires that the fermion gap should vanish near the critical point
between gapless and gapped phases. However, away from the critical
point in the insulating phase, the fermion gap obtained using
different polarization functions are no longer the same. The
momentum dependence of fermion gap at $T=0$ is shown in
Fig.\ref{fig:GapZeroT} for different values of $\lambda$. The solid
(dashed) lines are the results obtained using perturbative
(self-consistent) polarization function. The blue, red, and green
lines correspond to $1/\lambda = 0.1, 0.2, 0.3$ respectively. From
both Fig.\ref{fig:GapPSF} and Fig.\ref{fig:GapZeroT}, it is easy to
see that the dynamical fermion gap is significantly enhanced in the
Eliashberg formalism. Specifically, for graphene suspended in the
vacuum with $1/\lambda = 0.295$ \cite{CastroNetoPhys}, the fermion
gap calculated by the Eliashberg theory is greater by two orders of
magnitude than that calculated using perturbative polarization.

It is not hard to include the effects of temperature and chemical
potential. The dynamical fermion gap at finite $T$ and zero $\mu$ is
shown in Fig.\ref{fig:GapFinitT}, and the fermion gap at finite
$\mu$ with $T/\Lambda = 10^{-7}$ is shown in
Fig.\ref{fig:GapFinitMu}. In both of these diagrams, the solid
(dashed) lines are the results obtained using pertubative
(self-consistent) polarization function. The interaction strength is
fixed at $1/\lambda = 0.2$. The blue, red, green, and black lines
are the results at $T/\Lambda = 10^{-6}, 5\times10^{-6}, 10^{-5},
2\times10^{-5}$ respectively in Fig.\ref{fig:GapFinitT} and are the
results at $\mu/\Lambda = 10^{-6}, 5\times10^{-6}, 10^{-5},
2\times10^{-5}$ respectively in Fig.\ref{fig:GapFinitMu}. Here,
$\Lambda$ is the ultraviolet cutoff for momenta. Apparently, there
are substantial enhancement effects of dynamical fermion gap in both
cases of finite temperature and finite chemical potential. In the
Eliashberg theory, the static screening effects caused by finite
temperature and finite chemical potential are suppressed, so the
fermion gap is nearly independent of $T$ (see
Fig.\ref{fig:GapFinitT}) and $\mu$ (see Fig.\ref{fig:GapFinitMu}).

The fermion gap enhancement can be understood as follows. It is
well-known that the screening of Coulomb interaction is determined
by the density of states of fermions. Once the feedback of dynamical
fermion gap is taken into account, the density of states of fermions
is significantly reduced and the static or dynamical screening
effect becomes less important. As a consequence, the effective
Coulomb interaction becomes stronger, which in turn leads to larger
fermion gap.

In summary, we have studied the Eliashberg theory of excitonic phase
transition in graphene. After solving the coupled equations for
dynamical fermion gap and polarization function in a self-consistent
manner, we found a significant enhancement of dynamical fermion gap
in the excitonic insulating phase. The enhancement found within
Eliashberg formalism is owing to the suppression of static or
dynamical screening of Coulomb interaction by dynamical fermion gap.
Therefore, the self-consistent treatment of polarization function
should be used in the accurate computation of the dynamically
generated fermion gap. The validity of the Eliashberg theory is
justified by showing that the vertex corrections are suppressed by a
small factor in the large $N$ limit.

This work is supported by the National Natural Science Foundation of
China under grant No.11074234. G.Z.L. is also supported by the
Project Sponsored by the Overseas Academic Training Funds of
University of Science and Technology of China.

\section*{Appendix}

The polarization function $\Pi_{\mathrm{v}}(q)$ shown in Fig.2 is
defined as
\begin{eqnarray}
\Pi_{\mathrm{v}}(q)
&=&-N\int\frac{d^3k}{(2\pi)^3}\int\frac{d^3p}{(2\pi)^3}
\mathrm{Tr}\left[\gamma_{0}G_{0}(k+q)\gamma_{0}G_{0}(k+p+q)\gamma_{0}\right.
\nonumber \\
&&\left.\times G_{0}(k+p)\gamma_{0}D(p)G_{0}(k)\right].
\end{eqnarray}
To calculate this function, we will follow the method of Franz
\emph{et} \emph{al.} \cite{Franz03}. We are mainly interested in the
leading behavior of $\Pi_{\mathrm{v}}(q)$ in the $q \rightarrow 0$
limit. In this limit, the above integral has singularities as
$k\rightarrow 0$ and $k\rightarrow -p$. Thus, we may evaluate the
whole integral by expanding the regular parts of the integrand near
these two singular points. Keeping only the leading terms, we have
\begin{eqnarray}
\Pi_{\mathrm{v}}(q)&=&-N\int\frac{d^3k}{(2\pi)^3}\int\frac{d^3p}{(2\pi)^3}
\mathrm{Tr}\left[\gamma_{0}G_{0}(k+q)\gamma_{0}G_{0}(p+q)
\gamma_{0}\right.\nonumber \\
&& \left.\times G_{0}(p)\gamma_{0} D(p)G_{0}(k)\right] \nonumber \\
&& -N\int\frac{d^3k}{(2\pi)^3}\int\frac{d^3p}{(2\pi)^3}
\mathrm{Tr}\left[\gamma_{0}G_{0}(-p+q)
\gamma_{0}G_{0}(k+p+q)\gamma_{0}\right.\nonumber \\
&& \left.\times G_{0}(k+p)\gamma_{0}D(p)G_{0}(-p)\right]
\end{eqnarray}
Perform a variable shift, $k \rightarrow k-p$, for the second term,
then
\begin{eqnarray}
\Pi_{\mathrm{v}}(q) &=& 2N\mathrm{Tr}
\left[X(q)\int\frac{d^3k}{(2\pi)^3}
G_{0}(k)\gamma_{0}G_{0}(k+q)\right], \label{eqn:VertexPolarization}
\end{eqnarray}
where
\begin{eqnarray}
X(q)=-\int\frac{d^3p}{(2\pi)^3}
\gamma_{0}G_{0}(p+q)\gamma_{0}G_{0}(p)\gamma_{0}D(p).
\end{eqnarray}
The most leading term is found to be
\begin{eqnarray}
X(q) &=&-\frac{\gamma_{0}}{4\pi^2}\ln\left(\frac{\Lambda}{q}\right)
\left(\frac{e^2}{2\epsilon_{0}v_{F}}\right)\int_{0}^{\pi}d\theta
\frac{\cos^2\theta-\sin^2\theta}{1+\frac{Ne^2}{16\epsilon_{0}v_{F}}\sin\theta}.
\end{eqnarray}
In the large $N$ limit, it is possible to use the approximation
\begin{eqnarray}
\int_{0}^{\pi}d\theta \frac{\cos^2\theta -
\sin^2\theta}{1+\frac{Ne^2}{16\epsilon_{0}v_{F}}\sin\theta} \approx
\frac{32\epsilon_{0}v_{F}}{Ne^2}
\ln\left(N\frac{e^2}{8\epsilon_{0}v_{F}}\right),
\end{eqnarray}
so that
\begin{eqnarray}
X(q) &=& -\frac{4\gamma_{0}}{\pi^2}
\frac{\ln\left(N\frac{e^2}{8\epsilon_{0}v_{F}}\right)
\ln\left(\frac{\Lambda}{q}\right)}{N}. \label{eqn:Xq}
\end{eqnarray}
After substituting Eq.(\ref{eqn:Xq}) to
Eq.(\ref{eqn:VertexPolarization}), we finally get
\begin{eqnarray}
\Pi_{\mathrm{v}}(q) &=& -\frac{8}{\pi^2}
\frac{\ln\left(N\frac{e^2}{8\epsilon_{0}v_{F}}\right)
\ln\left(\frac{\Lambda}{q}\right)}{N} \Pi(q).
\end{eqnarray}

\end{document}